\documentclass[11pt]{article}
\usepackage{amsfonts}
 \usepackage{amssymb}
 \usepackage{amsmath}
 \newfont{\bbbold}{msbm10}

 \newfont{\goth}{eufm10 scaled \magstep1}

 \def\e{\epsilon}

 \def\m{\mu}

 \def\t{\tau}

 \def\be{\begin{equation}}\def\ee{\end{equation}}
 \def\bea{\begin{eqnarray}}\def\eea{\end{eqnarray}}
 \def\ba{\begin{array}}\def\ea{\end{array}}

 \def\del{\partial}

 \def\str{\rm str}

 \def\del{\partial}

 \def\3dt{\dot{3}}

 {}

 \def\bd{\begin{document}}
 \def\ed{\end{document}}
 \def\bea{\begin{eqnarray}}
 \def\ba{\begin{array}}\def\ea{\end{array}}
 \def\eea{\end{eqnarray}}
 \def\ft#1#2{{\textstyle{{\scriptstyle #1}\over {\scriptstyle #2}}}}
 \def\fft#1#2{{#1 \over #2}}
 \newcommand{\eq}[1]{(\ref{#1})}
 \def\eqs#1#2{(\ref{#1}-\ref{#2})}
 \def\det{{\rm det\,}}
 \def\tr{{\rm tr}}\def\Tr{{\rm Tr}}
  \def\str{{\rm str}} \def\diag{{\rm diag}}
 \def\sdet{{\rm sdet}}\def\symtr{{\rm symtr}}

\newcommand{\hoch}[1]{$^{#1}$}

\begin{document}

\thispagestyle{empty}


 \hfill{\today}

 \vspace{20pt}

\begin{center}
{\Large{\bf Reply to hep-th/0606265 }}
\vspace{30pt}

J. M. Drummond

\vspace{15pt}

Laboratoire d'Annecy-le-Vieux de Physique Th\'eorique LAPTH,

B.P. 110, F-74941 Annecy-le-Vieux, France

\vspace{60pt}

\end{center}

{\bf Abstract}
In hep-th/0411017 the Polchinski-Strominger effective string model was
examined and it was shown there that the spectrum of excitations is universal
up to and including terms of order $R^{-3}$ in the long distance
expansion. Subsequently the same 
result was claimed in hep-th/0606265 where certain criticisms of the earlier
work were made. In this note we demonstrate that the criticisms are wrong and
the methods and results of the earlier work are perfectly correct. In
particular we address the issue of higher order corrections to the action and
show that they were correctly given already in hep-th/0411017.

{\vfill\leftline{}\vfill \vskip  10pt

\baselineskip=15pt \pagebreak \setcounter{page}{1}

\section{Discussion of criticisms}

In \cite{drummond04} the Polchinski-Strominger effective string model
\cite{ps91} was
discussed. It was shown that the model remains conformal with non-trivial
energy-momentum tensor at one order higher in the long distance expansion with
respect to the original work of Polchinski and Strominger. 
In \cite{dm06} (DM) criticisms of the methods of \cite{drummond04} were made.
It should be noted that the later work \cite{dm06} agrees completely
with \cite{drummond04} in its conclusions, namely that the spectrum of
fluctuations of the Polchinski-Strominger effective string model is universal
up to and including $O(R^{-3})$ in the long distance expansion. It is given by
the Nambu-Goto spectrum to this order.

One criticism made by \cite{dm06} is that the higher order corrections to the
action are incorrectly given in the original work \cite{drummond04} on the
universality of the spectrum. In the
earlier work it is claimed that the next corrections to the action are
$O(R^{-6})$ and an explicit basis of four such terms was given. The criticism
of \cite{dm06} is that there are corrections at lower orders and terms which
are superficially $O(R^{-4})$ and $O(R^{-5})$ are given. We will first show
explicitly that the basis of \cite{drummond04} is correct
and then demonstrate that the terms constructed in the later work are merely
total derivatives plus a linear combination of terms in this basis.

First the explicit construction. We refer the reader to the original work of
Polchinski and Strominger \cite{ps91} or to \cite{drummond04,dm06} for the
details of the model and the terms that can appear in the action.
The terms in the Polchinski-Strominger action are constructed from worldsheet
derivatives of $d$ scalar fields $X^\mu$. There are two types of derivatives,
$\del_+$ and $\del_-$ and it is understood that we expand about a long string
vacuum where each first derivative is $O(R)$ and each higher derivative is
$O(1)$. Terms proportional to the lowest order equations of motion $\del_+
\del_- X$ are discounted since these can always be removed by a field
redefinition. Terms proportional to the lowest order constraints $\del_- X
\cdot \del_- X$ and $\del_+ X \cdot \del_+ X$ are also discounted. Finally
total derivatives in the Lagrangian are discounted since these do not
contribute to the action. The terms are allowed to be non-polynomial in the
derivatives of the fields but they must always be well-defined about the long
string vacuum which implies that the denominator of any given term can only be
$Z^n$ where $Z=\del_+ X \cdot \del_- X$ and $n$ is an integer. 

We consider spacetime scalar and worldsheet (1,1) tensor terms constructed
according to the above 
rules. The denominator of any given term is $Z^n$ as described above. The
numerator can contain contracted scalar pairs dressed with derivatives. Thus
it is made of factors of the following three types,

\be
\del_+^p X \cdot \del_-^q X \text{ (type I)}, \hspace{20pt} \del_+^p X \cdot
\del_+^q X \text{ (type II)},
\hspace{20pt} \del_-^p X \cdot \del_-^q X \text{ (type III)}.
\ee
   
There can be no $X$ with mixed derivatives since these are
(possibly after integration by parts) proportional to the lowest order field
equations. Factors of type I must contain at least three derivatives
since if it had just two it would be $Z$ and this would just change the power
of $Z$ in the denominator. Factors of types II and III must contain at
least four derivatives since if they had just two they would be the lowest
order constraints $\del_+X \cdot \del_+ X$ or $\del_- X \cdot \del_- X$ and if
they had just three they could be written as derivatives of the lowest order
constraints $\del_+^2 X \cdot \del_+ X = \frac{1}{2} \del_+(\del_+X \cdot
\del_+ X)$ and similarly for $- \longleftrightarrow +$. The order of each
factor is therefore either $R$ or $1$ depending on whether it has an $X$ with
a single derivative or not.

We now classify all terms up to and including $O(R^{-6})$ by the power $n$ of
$Z$ in the denominator. We need 
$(n+1)$ $\del_+$ and $(n+1)$ $\del_-$ in the numerator, i.e $2n+2$ derivatives
in total. Each pair contains at
least three derivatives so the number of pairs $p$ obeys $p\leq
\frac{2n+2}{3}$.  

For $n=1$ we can have only $p=1$. The only pair is therefore of type I,
\be
\frac{1}{Z} \del^2_+ X \cdot \del^2_- X. \label{n1}
\ee

For $n=2$ we can have $p=1,2$. For $p=1$ the only possibility is
\be
\frac{1}{Z^2} \del^3_+ X \cdot \del^3_- X. \label{n2p1}
\ee

For $p=2$ each factor must have three derivatives. The only possibility is
therefore two pairs of type I,
\be
\frac{1}{Z^2} \del^2_+ X \cdot \del_- X \del_+ X \cdot \del^2_- X. \label{n2p2}
\ee

For $n=3$ we can have $p=1,2$. For $p=1$ the only possibility is
\be
\frac{1}{Z^3} \del^4_+ X \cdot \del^4_- X. \label{n3p1}
\ee

For $p=2$ we consider first the case where one factor is of type
II (type III). This contains all the $\del_+$ ($\del_-$) so the other
factor must be of type III (type II). Each factor can have the derivatives
distributed as (2,2) or as (3,1) on the two $X$s. Therefore there are four
terms, 
\begin{align}
&\frac{1}{Z^3} \del^3_+ X \cdot \del_+ X \del^3_- X \cdot \del_- X,
  \label{n3p2II-IIIa}\\ 
&\frac{1}{Z^3} \del^3_+ X \cdot \del_+ X \del^2_- X \cdot \del^2_- X,
  \label{n3p2II-IIIb}\\
&\frac{1}{Z^3} \del^2_+ X \cdot \del^2_+ X \del^3_- X \cdot \del_- X,
  \label{n3p2II-IIIc}\\
&\frac{1}{Z^3} \del^2_+ X \cdot \del^2_+ X \del^2_- X \cdot \del^2_-
  X. \label{n3p2II-IIId}  
\end{align}

For $n=3$ it remains to consider the case $p=2$ where both factors are type
I. The 8 derivatives can be distributed over the two factors as (5,3) or as
(4,4). For the (5,3) case the factor with 3 derivatives can have 2 $\del_+$ or
2 $\del_-$ giving two possibilities,

\begin{align}
&\frac{1}{Z^3} \del^3_+ X \cdot \del^2_- X \del_+ X \cdot \del^2_- X,
  \label{n3p2I-I5-3a}\\
&\frac{1}{Z^3} \del^2_+ X \cdot \del^3_- X \del^2_+ X \cdot \del_-
  X.\label{n3p2I-I5-3b} 
\end{align}

For the (4,4) case the factors can have the derivatives distributed evenly or
not, 

\begin{align}
&\frac{1}{Z^3} \del^3_+ X \cdot \del_- X \del_+ X \cdot \del^3_- X,
  \label{n3p2I-I4-4a}\\
&\frac{1}{Z^3} \del^2_+ X \cdot \del^2_- X \del^2_+ X \cdot \del^2_-
  X.\label{n3p2I-I4-4b} 
\end{align}

For $n=4$ we can have $p=1,2,3$ but only $p=2,3$ can be $O(R^{-6})$ or
below. For $p=2$, each factor must have one $X$ with a single derivative
otherwise the term will be higher order than $O(R^{-6})$. This leaves 8
derivatives for the 2 remaining $X$s, meaning that the highest power on any
one $X$ must be 4 or 5 (5 is the highest since there are only 5 derivatives of
each type). If the highest power is 5 there are 2 possibilities,

\begin{align}
&\frac{1}{Z^4} \del^5_+ X \cdot \del_- X \del^3_- X \cdot \del_- X,
  \label{n4p25a} \\
&\frac{1}{Z^4} \del_+ X \cdot \del^5_- X \del^3_+ X \cdot \del_+
  X. \label{n4p25b} 
\end{align}

If the highest power is 4 there are again 2 possibilities,

\begin{align}
&\frac{1}{Z^4} \del^4_+ X \cdot \del_+ X \del^4_- X \cdot \del_- X,
  \label{n4p24a} \\
&\frac{1}{Z^4} \del^4_+ X \cdot \del_- X \del_+ X \cdot \del^4_-
  X. \label{n4p24b}
\end{align}

For $p=3$ the 10 derivatives must be distributed on the 3 factors as
(4,3,3). This implies that at least two must be of type I and these can
therefore contain at most 4 derivatives of one kind. Therefore the remaining
factor must contain both kinds of derivatives, i.e. be of type I also. The
factor carrying 4 derivatives can have them distributed as (3,1) or (2,2) on
the $X$s. In
the (3,1) case we have,

\begin{align}
&\frac{1}{Z^4} \del^3_+ X \cdot \del_- X \del_+ X \cdot \del^2_- X \del_+ X
  \cdot \del^2_- X, \label{n4p33-1a} \\
&\frac{1}{Z^4} \del_+ X \cdot \del^3_- X \del^2_+ X \cdot \del_- X \del^2_+ X
  \cdot \del_- X. \label{n4p33-1b} 
\end{align}

In the (2,2) case we have

\be
\frac{1}{Z^4} \del^2_+ X \cdot \del^2_- X \del^2_+ X \cdot \del_- X \del_+ X
\cdot \del^2_- X. \label{n4p32-2}
\ee

Finally for $n=5$ we can have $p=1,2,3,4$ but only $p=4$ can be $O(R^{-6})$
which occurs when each of the four factors carry 3 derivatives and hence
all are type I,

\be
\frac{1}{Z^5}\del^2_+ X \cdot \del_- X \del^2_+ X \cdot \del_- X \del_+ X
\cdot \del^2_- X \del_+ X \cdot \del^2_- X. \label{n5p4}
\ee

We now examine the equivalences due to integration by parts among the terms we
have produced. The terms 
(\ref{n1},\ref{n2p1},\ref{n3p1}) can all be rewritten as removable terms
proportional to 
the lowest order field equations plus terms with higher $n$. For example
(\ref{n1}) is equivalent to (\ref{n2p2}) which is the 
Polchinski-Strominger term.
Similarly 
(\ref{n3p2II-IIIa},\ref{n3p2II-IIIb},\ref{n3p2II-IIIc},\ref{n3p2II-IIId}) are
all equivalent up to terms of higher $n$, as are
(\ref{n3p2I-I5-3a},\ref{n3p2I-I5-3b},\ref{n3p2I-I4-4a},\ref{n3p2I-I4-4b}). 
The terms (\ref{n4p25a},\ref{n4p25b}) can also be written as terms with higher
$n$ (and higher order than $R^{-6}$) after integration by parts on one of the
5th order derivatives and dropping removable terms. The terms
(\ref{n4p24a},\ref{n4p24b}) can be integrated by parts to obtain terms of
order $R^{-7}$ or higher after dropping removable terms. The terms
(\ref{n4p33-1a},\ref{n4p33-1b},\ref{n4p32-2}) are all equivalent up to terms
with higher $n$.

Thus the only independent terms up to $O(R^{-6})$ can be chosen to be the
Polchinski-Strominger term (\ref{n2p2}) and the terms
(\ref{n3p2II-IIId},\ref{n3p2I-I4-4b},\ref{n4p32-2},\ref{n5p4}). The last four
are the terms explicitly given in \cite{drummond04}. It can be seen that they
are all $O(R^{-6})$. 

We now address the terms constructed in \cite{dm06} which are claimed to be of
lower order. We show explicitly that they are in fact equivalent to the terms
of $O(R^{-6})$ after integration by parts. First we deal with what is called
the `dominant class' in \cite{dm06}. This corresponds to terms with
denominators of the form $Z^M$ and an even number, $2N$,
of $X$s in the numerator, half of which (i.e $N$) have only single derivatives
($+$ or $-$) acting on them. The $N$ remaining $X$s each have at least two
derivatives acting on them, meaning that the total number of derivatives in the
numerator is at least $N+2N=3N$. 
There are $2M$ derivatives in the denominator and
so there are $2M + 2$ derivatives in the numerator (we are looking for a (1,1)
worldsheet tensor). Hence we have $2M + 2 \geq 3N$.
DM then argue that the order of such a term is $R^{N-2M}$, i.e. it is no
greater than $R^{-2N + 2}$. They then claim that for $N=3$ this implies the
leading behaviour is at most $O(R^{-4})$. In fact this can be improved
immediately by noting that the inequality $2M + 2 \geq 3N$ implies $2M + 2
\geq 3N + 1$ when $N$ is odd because $N$ and $M$ are both integer. So that
even by the analysis of DM such a term in the case $N=3$ would only be at most
order $R^{-5}$. Let us write down such a term. There are 3 $X$s carrying
single derivatives in the numerator and 3 $X$s carrying higher
derivatives. The total number of derivatives is even so the minimum is 10
for which the power $M$ of $Z$ in the denominator is 4. In this case
the only possibilities are then easily seen to be
\begin{align}
&\frac{1}{Z^4} \del_+ X^{\m_1} \del_+ X^{\m_2} \del_+ X^{\m_3} \del^2_+
X^{\m_4} \del^3_- X^{\m_5} \del^2_- X^{\m_6},\label{DMdoma} \\
&\frac{1}{Z^4} \del_+ X^{\m_1} \del_+ X^{\m_2} \del_- X^{\m_3} \del^3_+
X^{\m_4} \del^2_- X^{\m_5} \del^2_- X^{\m_6} \label{DMdomb}
\end{align}
and the terms with $+ \longleftrightarrow -$. The indices $\m_i$ are to be
contracted with some combination of metric tensors. In the first case
(\ref{DMdoma}) there is no way to do this without producing the lowest order
constraint $\del_+ X \cdot \del_+ X$ or its derivative. In the second case
(\ref{DMdomb}) the only way which avoids 
producing a factor of $Z$ in the numerator or 
producing the lowest order constraints (or their derivatives) is to contract
the $\del_+ X$ factors with the $\del^2_- X$ factors and the $\del_- X$ with
the $\del^3_+ X$ and similarly for $+ \longleftrightarrow -$. This leaves the
terms 
\be
\frac{1}{Z^4} \del^3_+ X \cdot \del_- X \del_+ X \cdot \del^2_- X \del_+ X
\cdot \del^2_- X  \text{ and } + \longleftrightarrow -.
\ee
These are the terms (\ref{n4p33-1a},\ref{n4p33-1b}) from the previous
discussion. 
They are naively of
order $R^{-5}$ as would follow from the (improved version presented above of
the) analysis of 
DM. However as claimed above we can rewrite them after integration by parts (to
be completely explicit we give the whole calculation here):
\begin{align}
&\frac{1}{Z^4} \del^3_+ X \cdot \del_- X \del_+ X \cdot \del^2_- X \del_+ X
\cdot \del^2_- X \label{line1}\\
=&\del_+ \bigl( \frac{1}{Z^4} \del^2_+ X \cdot \del_- X \del_+ X \cdot
\del^2_- X \del_+ X \cdot \del^2_- X \bigr) \label{line2}\\
&-\frac{4}{Z^5} (\del^2_+ X \cdot \del_- X + \del_+ X \cdot \del_+ \del_-
X) \del^2_+ X \cdot \del_- X \del_+ X \cdot \del^2_- X \del_+ X \cdot \del^2_-
X \label{line3}\\
&-\frac{2}{Z^4} \del^2_+ X \cdot \del_- X (\del^2_+ X \cdot \del^2 _- X +
\del_+ X \cdot \del_+ \del^2_- X) \del_+ X \cdot \del^2_- X \label{line4} 
\end{align}
The total derivative term (line (\ref{line2})) does not contribute to the
action. The second term in the brackets in line (\ref{line3}) and the second
term in the brackets in line (\ref{line4}) both contain mixed derivatives
acting on one of the $X$s and hence can be removed by field
redefinitions. This leaves the the first term in the brackets from line
(\ref{line3}) and the first term in the brackets in line (\ref{line4}),
\begin{align}
&-\frac{4}{Z^5} \del^2_+ X \cdot \del_- X  \del^2_+ X \cdot \del_- X \del_+ X
 \cdot \del^2_- X \del_+ X \cdot \del^2_- X \notag \\
&-\frac{2}{Z^4} \del^2_+ X \cdot \del_- X \del^2_+ X \cdot \del^2 _- X \del_+
 X \cdot \del^2_- X. 
\end{align}
This is a linear combination of the terms (\ref{n4p32-2},\ref{n5p4}). 
They are order $R^{-6}$.

The same analysis holds for the term which DM call the `dominant case in the
subdominant class' and which they claim is order $R^{-5}$. The missing step
is again integration by parts. 
Thus we have shown that the terms given in \cite{drummond04} are correct and
the terms claimed by DM to be order $R^{-4}$ and order $R^{-5}$ are only
superficially so.  

Spacetime pseudoscalar terms were not classified in \cite{drummond04}. This is 
because their structure is dimension dependent and they can be consistently
dropped from the action and transformation law while retaining conformal
invariance up to the relevant order. However it was noted in \cite{drummond04}
that in four spacetime dimensions there is a term of order $R^{-2}$. It is
given by
\be
\frac{1}{Z^2} \del^2_+ X^{\m_1} \del_+ X^{\m_2} \del^2_- X^{\m_3} \del^2_+
X^{\m_4} \e_{\m_1 \m_2 \m_3 \m_4}. \label{pseudo}
\ee
This term is much like the Polchinski-Strominger term (the only difference
being that the indices are contracted with $\e_{\m_1 \m_2 \m_3 \m_4}$ instead
of $\eta_{\m_1 \m_4} \eta_{\m_2 \m_3}$ ). DM also discuss this term. They 
point out that, after expanding around the vacuum of winding number one,
\be
X^\m = R(e_+^\m \t^+ + e_-^\m \t^-) + Y^\m, \hspace{20pt} e_{\pm}^2 =
0, \hspace{15pt}  e_+ \cdot e_- = -\frac{1}{2}, \label{vac}
\ee
the term is
\begin{align}
&4\Bigl( \frac{1}{R^2} \del^2_+ Y^{\m_1} e_+^{\m_2} \del^2_- Y^{\m_3}
  e_-^{\m_4}  
+ \frac{1}{R^3} \del^2_+ Y^{\m_1} e_+^{\m_2} \del^2_- Y^{\m_3} \del_- Y^{\m_4}
+ \frac{1}{R^3} \del^2_+ Y^{\m_1} \del_+ Y^{\m_2} \del^2_- Y^{\m_3}
e_-^{\m_4} \notag \\
&+ \frac{4}{R^3} \del^2_+ Y^{\m_1} e_+^{\m_2} \del^2_- Y^{\m_3} e_-^{\m_4} (e_+
\cdot \del_- Y + \del_+ Y \cdot e_-) 
\Bigr) \e_{\m_1 \m_2 \m_3 \m_4} + O(R^{-4})
\end{align}
Each of the order $R^{-2}$ terms and order $R^{-3}$ terms is proportional to
$\del_+ \del_- Y$ after integration by parts and hence these terms can all be
removed by field redefinitions. Hence, after expanding around the vacuum
(\ref{vac}), the above term is actually of order $R^{-4}$. We stress that
it is due to field redefinitions of the form $\hat{Y} = Y + ...$ that the
order $R^{-2}$ and $R^{-3}$ terms can be removed. 

Exactly the same analysis holds for the Polchinski-Strominger term
(one just replaces $\e_{\m_1 \m_2 \m_3 \m_4}$ with $\eta_{\m_1 \m_4}
\eta_{\m_2 \m_3}$ in the above analysis).  Indeed this is precisely the
reasoning which leads to the field redefinition given by equation (2.19) of 
\cite{drummond04}. Thus, after expanding around the vacuum (\ref{vac}), the
Polchinski-Strominger term is also order $R^{-4}$ (as shown explicitly in
\cite{drummond04}). 

From the fact that the term (\ref{pseudo}) is order $R^{-4}$ in the long
string vacuum, DM conclude that it can be dropped. This is clearly not
enough. The 
same argument could be applied to the Polchinski-Strominger term and this
clearly cannot be dropped. The reason is that the Polchinski-Strominger term
induces terms in the conformal transformation law which are order $R^{-1}$,
even in terms of the redefined field $\hat Y$. Therefore, even though it is
order $R^{-4}$ in the long string vacuum, it has an effect on the energy-
momentum tensor at order $R^{-1}$. To see the expression for the energy-
momentum tensor in terms of $\hat{Y}$ (equation (2.20) of \cite{drummond04})
one merely has to invert the expression for $\hat Y$ (equation (2.19) 
of \cite{drummond04}). The expression for $Y$ in terms of $\hat Y$ is
precisely the same as equation (2.19) after swapping $Y$ and $\hat Y$ and
swapping the sign of the higher order terms.
Then one substitutes into the expression for the
energy-momentum tensor given in terms of $Y$ (equation (2.16) of
\cite{drummond04}), which itself is derived from the expression in terms of
$X$ (equations (2.11) and (2.12) of \cite{drummond04}). To properly address
the term (\ref{pseudo}) one has to consider potential corrections to the
conformal transformation law and the effect they may have on the
energy-momentum tensor. It is clear however that such a term cannot
effect the spectrum however since $p^2$ is still a spacetime scalar up
to the relevant order.

DM criticise the use of the field redefinition in \cite{drummond04} which we
have described again above on the grounds that it is unclear how it leads to
the expression (2.20) of \cite{drummond04} for the energy-momentum tensor. As
we have seen this objection is spurious - one merely has to make the simple
substitutions just described. In fact this procedure is exactly equivalent to
the procedure outlined by DM when they iteratively solve the equation of
motion to obtain an expression for $T_{--}$ in terms of what they call
$Y_0$. One can see explicitly that expression (19) of DM \cite{dm06} is
exactly the same as expression (2.20) of \cite{drummond04} after the
identification of $Y_0$ and $\hat Y$ (which both obey the lowest order equation
of motion up to the relevant order). 
DM also claim that they use no such field redefinitions themselves.  
In fact, as we have just seen, they use
precisely the {\it same} field redefinition themselves to argue that the term
(\ref{pseudo}) is order $R^{-4}$ in the long string vacuum. 
Finally, we should note that questions of changes in the path integral measure
are unnecessary in the case of the field redefinition in
\cite{drummond04}. What we have shown in \cite{drummond04} is that, as an
expansion about the long string vacuum, the theory remains conformal with
critical central charge and non-trivial energy-momentum tensor at one order
higher in $R^{-1}$ with respect to the original work \cite{ps91}. It
should be pointed out that DM do not bother to verify the operator product of
the energy-momentum tensor and hence do not bother to verify that the
expression for the Virasoro generators really does satisfy the Virasoro algebra
up to the next order.

\section{Conclusions}

In this note we have addressed the criticisms made in \cite{dm06} of the
earlier work \cite{drummond04}. We have shown:

1. The classification of higher order terms in the action presented in
   \cite{drummond04} is correct and the relevant terms are order
   $R^{-6}$. 

2. The terms presented by DM in \cite{dm06} have explicitly been
   shown to be total derivatives plus a linear combination of terms of the
   basis presented in \cite{drummond04}. Thus they are only superficially of
   lower order despite the claims of DM.

3. The claim of DM that they use no field redefinitions in their analysis is
   false. In fact they use exactly the {\it same} redefinition already
   discussed in \cite{drummond04} when they discuss the pseudoscalar term
   (\ref{pseudo}). Their claim that the field redefinition obscures the
   derivation of the energy momentum-tensor in \cite{drummond04} is spurious -
   one simply rewrites expression (2.16) of \cite{drummond04} in terms of
   $\hat Y$. 

4. DM imply that corrections of order $R^{-4}$ in the action can simply be
   ignored. As we have seen, it is also necessary to address corrections to
   the transformation law (as in the case of the Polchinski-Strominger
   term). One has to show that there are no relevant corrections to
   the transformation law, a fact which follows immediately 
   when one realises that the next relevant corrections to the action
   are in fact order $R^{-6}$.

We therefore conclude that the work \cite{dm06} is, at best, merely a partial 
reconstruction of the results of the earlier work \cite{drummond04}. In
addition we have explicitly demonstrated that it contains technical and
conceptual errors which render its criticisms of the earlier work invalid.

\end{document}